\documentclass[10pt,a4paper]{article}
\usepackage{amsmath}
\usepackage{amsfonts}
\usepackage{amssymb}
\usepackage{graphicx}
\usepackage{textcomp}
\usepackage{url}
\usepackage{lipsum}
\usepackage{setspace}
\usepackage{comment}
\usepackage{color}
\usepackage[numbers,sort&compress]{natbib}
\usepackage[usenames,dvipsnames,svgnames,table]{xcolor}
\usepackage[lofdepth,lotdepth,caption=false]{subfig}
\usepackage[pdfstartview=FitH]{hyperref}
\usepackage{hyperref}
\newsavebox{\bigleftbox}
\hypersetup{
 colorlinks,
 citecolor=Blue,
 linkcolor=Blue,
 urlcolor=Blue
}

\makeatletter
 \def\footnoterule{\kern-3\p@
   \noindent\hrulefill \kern 2.8\p@} 
\makeatother
\usepackage[left=3.00cm, right=3.00cm, top=2.00cm, bottom=2.00cm]{geometry}

\usepackage{fancyhdr}

\title{\textbf{Predicting $\Psi$-BN: computational insights into its mechanical, electronic, and optical characteristics}}
\author{
	F. F. Monteiro $^{1,2,\dag}$,
    K. A. L. Lima$^{1,2,\P}$, 
	L. A. Ribeiro Junior$^{1,2,\S}$
	}
\date{}

\begin{document}
    \maketitle
	\vspace{-0.6cm}
	\begin{center}\small
	\textit{Institute of Physics, University of Bras\'ilia, 70910-900, Bras\'ilia, Brazil}\\
            \textit{Computational Materials Laboratory, LCCMat, Institute of Physics, University of Bras\'ilia, 70910-900, Bras\'ilia, Brazil}\\
		\phantom{.}\\ \hfill
		$^{\dag}$\url{fmonteiro@unb.br}\hfill
        $^{\P}$\url{kleutonantuneslopes@prp.uespi.br}\hfill
		$^{\S}$\url{ribeirojr@unb.br}\hfill
		\phantom{.}
	\end{center}
	

\onehalfspace

\noindent\textbf{Abstract: Computational materials are pivotal in understanding various material classes and their properties. They offer valuable insights for predicting novel structures and complementing experimental approaches. In this context, PSI ($\Psi$)-graphene has recently emerged as a stable two-dimensional carbon allotrope composed of 5-6-7 carbon rings. Using Density Functional Theory (DFT) calculations, we explored its boron nitride counterpart's mechanical, electronic, and optical characteristics, $\Psi$-BN. Our findings reveal that $\Psi$-BN possesses a band gap of 4.59 eV, as determined at the HSE06 level. Phonon calculations and ab initio molecular dynamics simulations demonstrate its structural and dynamic stability. $\Psi$-BN exhibits a low formation energy of -7.48 eV. Moreover, we observe strong ultraviolet (UV) activity in $\Psi$-BN, implying its potential as an efficient UV collector. We have also determined its critical mechanical properties, including elastic stiffness constants, Young's modulus (ranging from 250 to 300 GPa), and a Poisson ratio of 0.7, collectively providing valuable insights into its mechanical behavior.}

\section{Introduction}

The hexagonal boron nitride \cite{shi2010synthesis} (h-BN) lattice, composed of boron and nitrogen atoms arranged in a structure reminiscent of graphene \cite{geim2009graphene}, presents a wide range of properties that hold significant implications for optoelectronic applications \cite{wang2017electrical}. These distinctive characteristics encompass exceptional thermal insulation \cite{an2021quasi}, robust dielectric properties \cite{laturia2018dielectric}, and mechanical resilience \cite{yang2021intrinsic}. The h-BN has a diverse array of uses, such as electronic substrates \cite{sachs2011adhesion}, protective coatings \cite{husain2013marine}, solid lubricants \cite{kimura1999boron}, and optical components \cite{zunger1976optical}. These features position h-BN as an intriguing and versatile candidate among two-dimensional materials \cite{zhang2017two}. The achievements of h-BN in these applications have ignited numerous investigations into the design of two-dimensional computational materials incorporating boron and nitrogen elements \cite{shahrokhi2017new,revabhai2023progress,arenal2015boron,roy2021structure,ihsanullah2021boron,amiri2020electro,takahashi2017structural,pontes2021electronic,ma2022bn,monteiro2023mechanical}.

Recent efforts in material synthesis have yielded a range of fascinating 2D carbon allotropes, including well-known examples such as graphene \cite{geim2009graphene}, the biphenylene network \cite{fan2021biphenylene}, $\gamma$-graphyne \cite{desyatkin2022scalable}, and the monolayer fullerene network \cite{hou2022synthesis}. These discoveries have not only spurred theoretical proposals for analogous structures composed of boron and nitrogen elements \cite{tawfik2017first,shahrokhi2017new,asadpour2015mechanical,yadav2023bn} but have also led to explorations into the world of 2D carbon structures, even though they have not yet been experimentally realized. Prominent examples of this endeavor also involved boron and nitrogen atoms as constituents, yielding materials such as penta-graphene \cite{zhang2015penta} (penta-BN \cite{amiri2020electro}), phagraphene \cite{wang2015phagraphene} (pha-BN \cite{pontes2021electronic}), and T-carbon \cite{sheng2011t} (T-BN \cite{takahashi2017structural}).

Computational methods are pivotal in predicting new 2D nanomaterials, offering invaluable insights. This predictive process assesses the feasibility of experimental synthesis and provides a preliminary evaluation of their properties and potential applications. In the context of boron nitride materials, computational anticipation of novel candidates holds particular significance. This trend is due to boron nitride allotropes' diverse structures and unique properties. It can enrich our understanding of the various options for advanced computational materials applications.

One noteworthy computational carbon allotrope garnered significant attention is $\Psi$-graphene \cite{li2017psi}. This material consists of a single, atom-thick 2D monolayer composed entirely of carbon atoms. Its unique structure is formed by organizing carbon skeletons reminiscent of s-indacenes, a small hydrocarbon molecule with the chemical formula C$_{12}$H$_8$. Combining density functional theory (DFT) calculations with ab initio molecular dynamics (AIMD) simulations, Li \textit{et al.} have uncovered intriguing properties of $\Psi$-graphene \cite{li2017psi}. It exhibits exceptional dynamic and thermal stability, surpassing the mechanical rigidity of penta-graphene and h-BN monolayers. Additionally, $\Psi$-graphene demonstrates metallic properties under ambient conditions \cite{li2017psi}.

Due to its unique structure and inherent metallic properties, $\Psi$-graphene displays a shallow energy barrier for diffusion and a robust capacity for storing lithium ions \cite{li2017psi}. These features make $\Psi$-graphene a promising candidate as an anodic material for lithium-ion batteries. It is worth noting that it boasts the lowest energy state among all reported 2D carbon allotropes featuring 5-6-7 carbon ring formations. Given these remarkable characteristics, investigating the electronic and structural properties of $\Psi$-graphene's counterpart, boron nitride ($\Psi$-BN), could significantly enhance our understanding of the potential impact of these materials in various optoelectronic applications.

This study explores the mechanical, electronic, and optical properties of $\Psi$-BN, a material formed from boron and nitrogen atoms and serving as a counterpart to $\Psi$-graphene. Our investigation relies on the application of DFT calculations and AIMD simulations. We gain valuable insights into this intriguing material's electronic and structural characteristics using our computational approach. A central focus of our research is the assessment of its structural stability. Notably, the $\Psi$-graphene framework has shown potential in guiding the development of robust BN-based materials with similar structural attributes.

\section{Methodology}

\begin{figure}[!htb]
    \centering
    \includegraphics[width=0.6\linewidth]{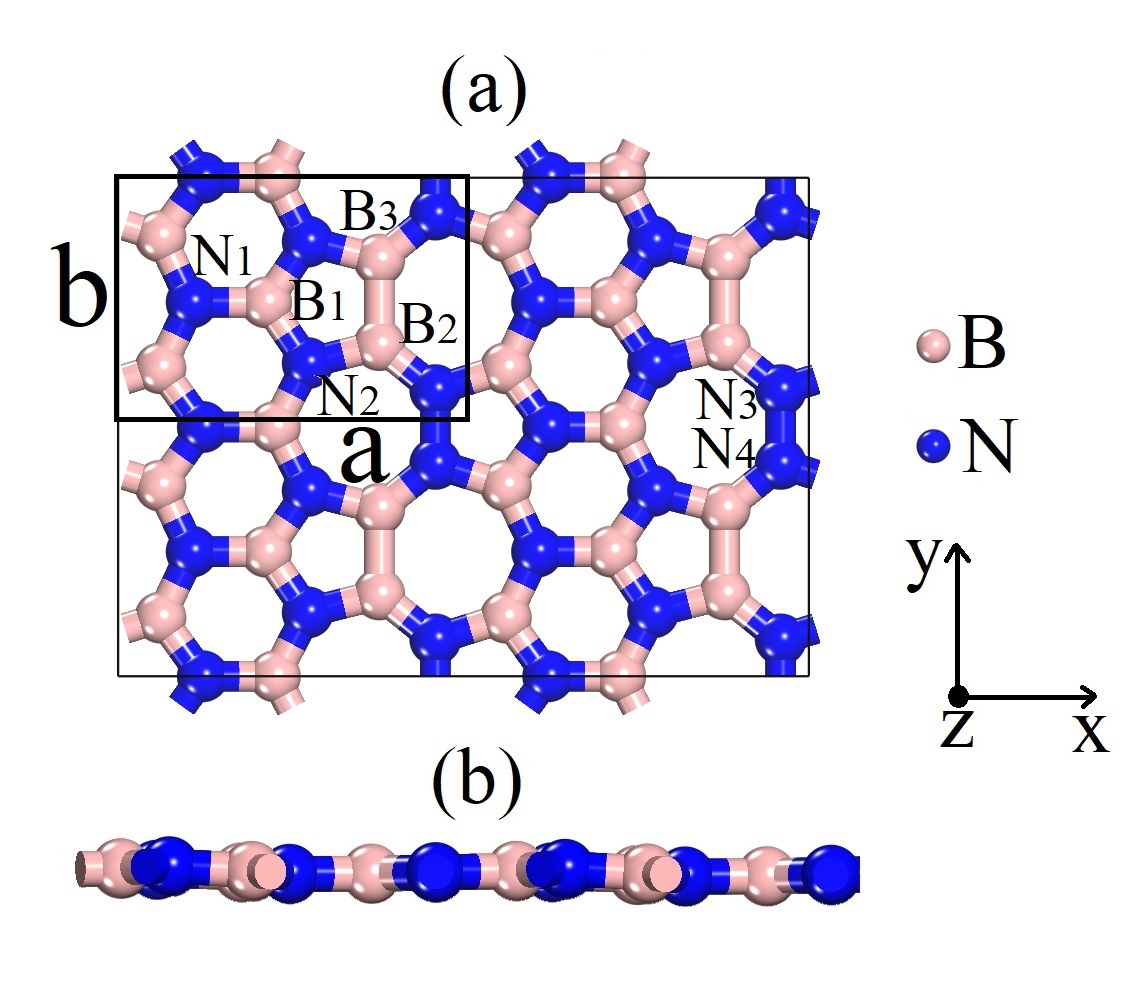}
    \caption{Schematic representation of the $\Psi$-BN monolayer in the (a) top and (b) side views. The black rectangle defined by lattice vectors $a$ and $b$ highlights the unit cell. Pink and blue represent boron and nitrogen atoms. The N and B labels show atoms establishing distinct B-N bonds.}
    \label{fig:sys}
\end{figure}

We employed DFT simulations to investigate the mechanical, electronic, and optical properties of $\Psi$-BN (illustrated in Figure \ref{fig:sys}) using the CASTEP \cite{clark2005first} code. In our calculations, we applied the generalized gradient approximation (GGA) for exchange and correlation functions, explicitly using the Perdew-Burke-Ernzerhof (PBE) \cite{perdew1996generalized} functional and the Heyd-Scuseria-Ernzerhof (HSE06) \cite{heyd2003hybrid} hybrid functional. To ensure accuracy in modeling the interactions between nuclear electrons of each atomic species, we utilized norm-conserving pseudopotentials designed explicitly for boron and nitrogen, as provided within the CASTEP framework.

The Broyden–Fletcher–Goldfarb–Shannon (BFGS) unrestricted nonlinear iterative algorithm was used to achieve electronic self-consistency \cite{head1985broyden,PFROMMER1997233}. Our calculations also employed a plane-wave basis set with an energy cutoff of 700 eV, and we established a convergence criterion for energy at 1.0 $\times 10^{-5}$ eV. To fully relax the $\Psi$-BN lattice, we applied periodic boundary conditions, ensuring that the residual forces on each atom remained below 1.0 $\times 10^{-3}$ eV/\r{A}, and the pressure remained below 0.01 GPa. 

We fixed the basis vector in the z direction during the lattice optimization while employing a k-grid of $10\times10\times1$. Additionally, for electronic and optical computations, we adopted k-grids of $15\times15\times1$ and $5\times5\times1$ for the PBE and HSE06 approaches, respectively. Partial Density of States (PDOS) was determined at the HSE06 level, using a k-grid of $20\times20\times1$. Elastic properties were derived using the LDA/CA-PZ method \cite{PhysRevLett.45.566,PhysRevB.23.5048}. A vacuum region spanning $20$ \r{A} was settled to minimize artificial interactions among periodic images.

A lower k-grid resolution ($5\times5\times1$) for our HSE06 calculations was used instead of the k-grid used in the PBE approach. Given the significant computational cost associated with the HSE06 method, this decision primarily arises from practical computational considerations. We conducted thorough convergence tests to ensure the adequacy of our chosen k-grid resolution for HSE06 calculations. Our analysis established that the selected k-grid resolution maintains the required precision and reliability within an acceptable margin of error.

For phonon calculations, we employed the linear response methodology with a grid spacing of 0.05~1/\r{A} and a convergence tolerance of 10$^{-5}$~eV/\r{A}$^2$. To evaluate the mechanical properties of the $\Psi$-BN lattice, we utilized the stress–strain method based on the Voigt–Reuss–Hill approximation \cite{Zuo:gl0256,10.1063/1.1709944}. Additionally, we calculated the effective mass ($m^*$) for electrons and holes, a fundamental parameter in determining the overall carrier mobility in 2D materials. The calculation involved fitting the band dispersion to

\begin{equation}
 \displaystyle m^{*}=\hslash^2\left(\frac{\partial^2E(k)}{\partial k^2}\right)^{-1}.   
\end{equation}

\section{Results}

Figure \ref{fig:sys} presents the optimized lattice arrangement of $\Psi$-BN. The geometry optimization calculations yield similar lattice parameters using the PBE and HSE06 methods. For this discussion, we emphasize the results obtained with the HSE06 method, which are visualized in Figure \ref{fig:sys} and detailed in Table \ref{tab:bonds}. As seen in Figure \ref{fig:sys}, B-B and N-N bonds are noticeable within the seven-atom rings, with a similar bonding pattern in the five-atom rings. In contrast, hexagons consistently exhibit B-N bonds. The lattice vectors $a$ and $b$ measure 6.67 \r{A} and 4.82 \r{A}, respectively (see Figure \ref{fig:sys}). The crystal structure of $\Psi$-BN conforms to an orthorhombic arrangement explicitly aligned with the PM (CS-1) space group.

\begin{table}[!htb]
\caption{Bond distances for the atoms highlighted in Figure \ref{fig:sys} calculated at the HSE06 level.}
\centering
\label{tab:bonds}
\begin{tabular}{|c|c|c|c|}
\hline
Bond Type & Bond Length (\r{A}) & Bond Type & Bond Length (\r{A}) \\ \hline
N1-B1    & 1.51                          & B2-B3    & 1.68                          \\ \hline
N2-B2    & 1.41                        &   B2-N3    & 1.49                       \\ \hline
B1-N2    & 1.47                         & N3-N4    & 1.36                         \\ \hline
\end{tabular}
\end{table}

\begin{figure}[!htb]
	\centering
	\includegraphics[width=0.6\linewidth]{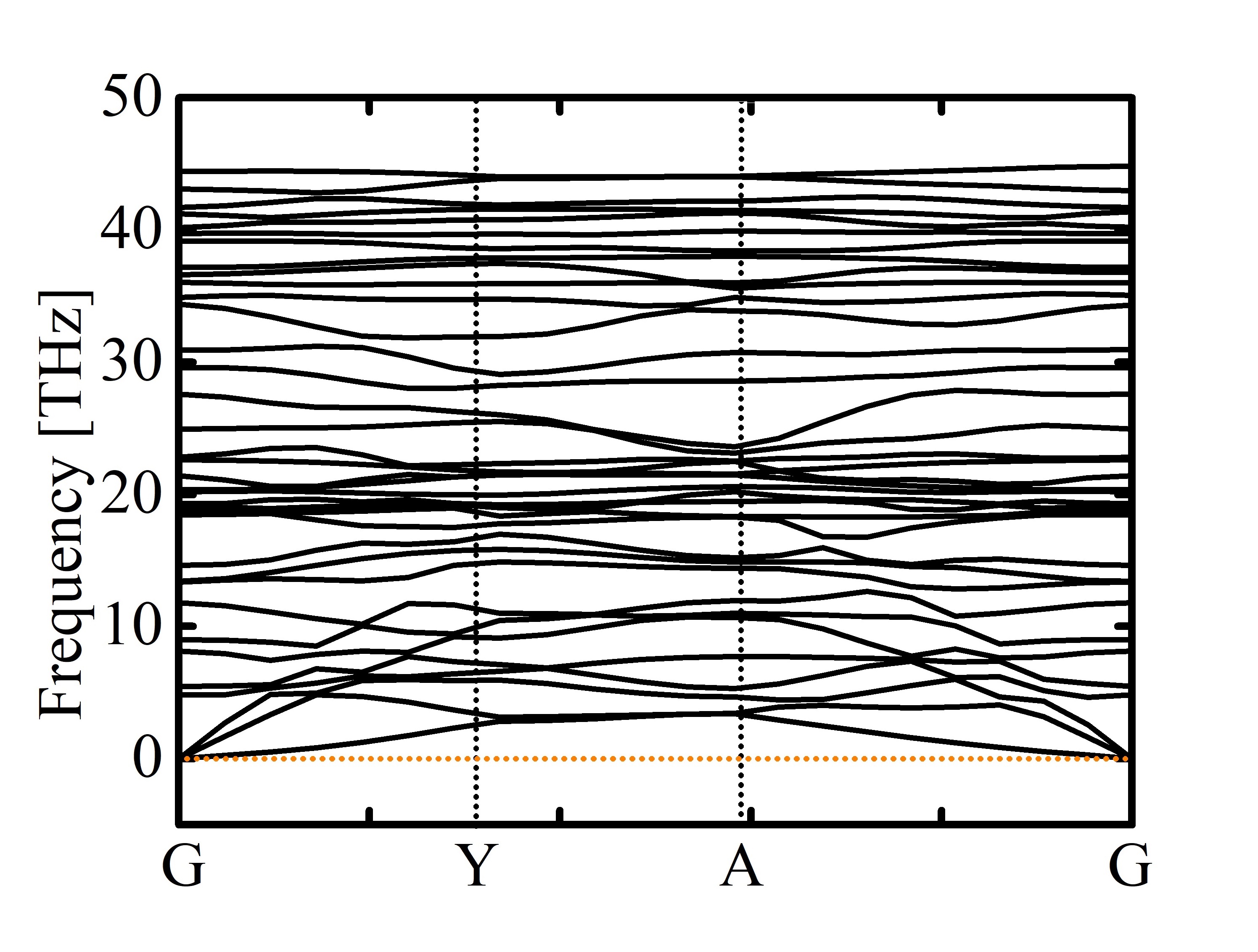}
	\caption{Phonon band structure of $\Psi$-BN calculated at PBE level.}
	\label{fig-phonons}
\end{figure}

The phonon dispersion of the $\Psi$-BN lattice is illustrated in Figure \ref{fig-phonons}. One can observe the absence of imaginary frequencies, suggesting the dynamical stability of this material. This feature represents one of the crucial indicators affirming the stability of $\Psi$-BN. Additionally, the lack of a band gap between acoustic and optical modes implies a significant scattering rate and short phonon lifetimes, contributing to the material's relatively modest lattice thermal conductivity. Furthermore, our analysis revealed a formation energy value of -7.48 eV for the $\Psi$-BN lattice.

Figure \ref{fig-phonons} also reveals an overlap between the acoustic and optical phonon bands for  $\Psi$-BN. This overlap carries substantial implications for the material's physical properties. First, it indicates the absence of a distinct band gap between the acoustic and optical modes. One notable consequence is its impact on thermal conductivity. The overlap between the acoustic and optical phonon bands suggests a significant scattering rate and a short phonon lifetime within the material. Phonons are the primary heat carriers in crystals, and their scattering is directly linked to thermal conductivity. In materials with such an overlap, phonons can scatter more easily, leading to reduced phonon mean accessible paths and lower thermal conductivity. This property is particularly advantageous for applications where thermal insulation or thermoelectric materials with low thermal conductivity are desired.

The overlap between the acoustic and optical phonon bands can also influence the material's mechanical properties. As mentioned above, the absence of a well-defined band gap implies a relatively high degree of phonon delocalization within the material. Consequently, the material can undergo structural deformations more quickly when subjected to mechanical stress. This characteristic can impact the material's elastic properties and response to strain. 

The distinct $\Psi$-BN optical properties, discussed below, are also impacted by the overlap between the acoustic and optical phonon bands. It suggests that the material can efficiently scatter and absorb photons within the frequency range corresponding to the overlap. These features can lead to enhanced optical absorption in this region, a characteristic often observed in semiconductors with a pronounced scattering rate. In this way, the material may exhibit unique optical behaviors, making it suitable for applications in photonics and optoelectronics, such as photodetectors.

\begin{figure}[!htb]
	\centering
	\includegraphics[width=0.6\linewidth]{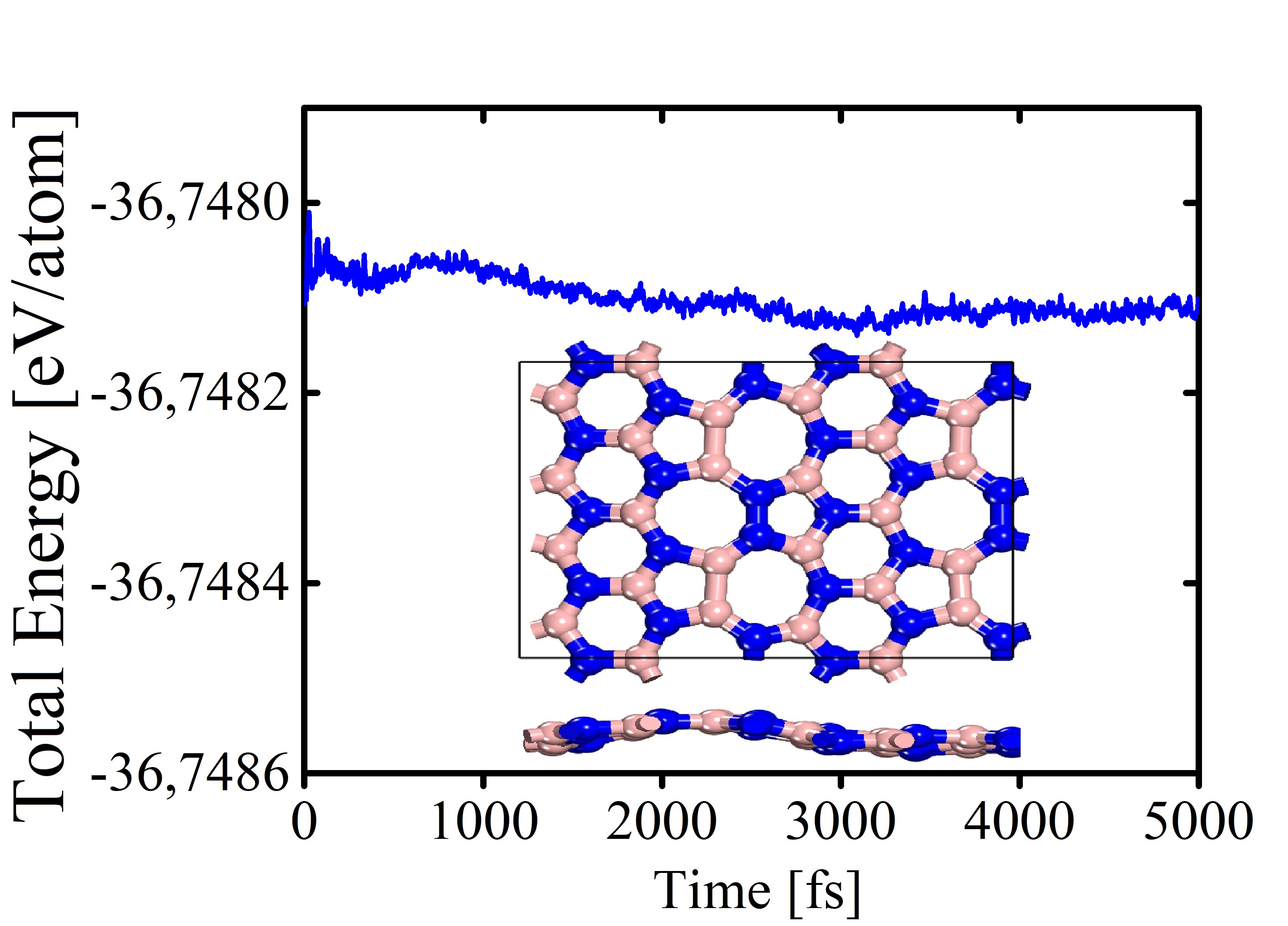}
	\caption{Time evolution of the total energy per atom in the $\Psi$-BN lattice at 1000K, using the PBE approach. The insets show the top and side views of the final AIMD snapshot at 5000 fs.}
	\label{fig-aimd}
\end{figure}

We also conducted AIMD simulations to assess the dynamic and thermal stability of $\Psi$-BN. During these simulations, we tracked the time evolution of the total energy per atom over 5000 fs of simulation, exposing the material to a temperature of 1000 K, as illustrated in Figure \ref{fig-aimd}.

In a thermally stable system, the total energy per atom should remain relatively constant, indicating an even energy distribution within the material. For our AIMD simulations, we employed a $2\times2\times1$ supercell containing 48 atoms. As depicted in Figure \ref{fig-aimd}, the temporal evolution of the total energy exhibits a nearly flat pattern with minimal fluctuations. The MD panels within the figure reveal deviations in planarity and bond distances among the lattice atoms due to the elevated temperature. However, no bond breakages or reconstructions occur. The lattice topology at 1000 K remains consistent with the optimized structure (see Figure \ref{fig:sys}). These observations collectively affirm the robust thermal stability of $\Psi$-BN.

\begin{figure}[!htb]
	\centering
	\includegraphics[width=\linewidth]{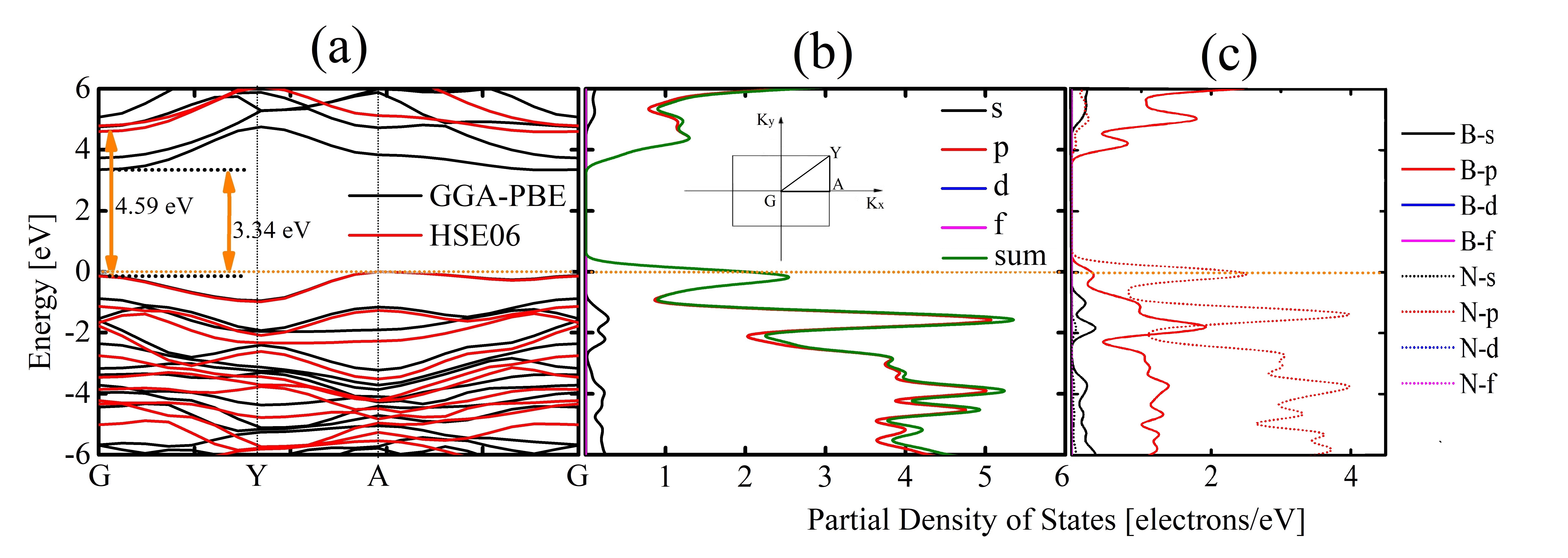}
	\caption{(a) Electronic band structure and (b) partial density of states (PDOS) for the $\Psi$-BN monolayer. The band structures were obtained using the PBE (black) and HSE06 (red) approaches. PDOS was calculated at the HSE06 level. In panel (b), the inset shows the integration path.}
	\label{fig-bands}
\end{figure}

We now delve into the band structure characteristics of the $\Psi$-BN lattice. In Figure \ref{fig-bands}(a), we present the band structure profiles obtained using the PBE (black) and HSE06 (red) methods. The calculated band gaps are 3.34 eV and 4.59 eV for the PBE and HSE06 levels, respectively.

It is important to note that PBE calculations tend to underestimate band gaps. Therefore, we employ the hybrid functional HSE06 to more accurately determine the electronic and optical properties of $\Psi$-BN. The PBE and HSE06 band structures define $\Psi$-BN as a wide band gap semiconductor, with energy bands near the Fermi level displaying dispersion. This dispersion signifies the delocalized nature of electronic states within this material.

For context, h-BN exhibits an insulating band gap of approximately 5.95 eV \cite{cassabois2016hexagonal}. The smaller band gap observed in $\Psi$-BN compared to h-BN can be attributed to its ring topology and the presence of B-B and N-N bonds in its lattice. These elements introduce unique pathways for electronic transport that are absent in h-BN. 

It is worth mentioning that the disparities in predicting the band gap values for $\Psi$-BN stem from the computational methods' inherent limitations and approximations. The HSE06 method, as a hybrid functional, introduces a portion of exact exchange alongside the exchange-correlation functional. Incorporating non-local exchange interactions enhances the accuracy of electronic structure predictions, particularly for systems characterized by significant electron localization or delocalization, such as semiconductors. Consequently, the band gap calculated using HSE06 is often considered to provide a more reliable representation of electronic properties. 

The partial density of states (PDOS) of Psi-BN, using the HSE06 hybrid functional, is illustrated in Figure \ref{fig-bands}(b). A noticeable trend emerges, revealing a pronounced dominance of the p-states. This outcome signifies that the engagement of p-orbitals primarily drives electronic transitions and interactions. Such orbitals are commonly linked to directional bonding phenomena. 

Figure \ref{fig-bands}(b) comprehensively overviews the contributions of s, p, d, and f orbitals from both atomic species within each lattice. In Figure \ref{fig-bands}(c), we delve deeper into the PDOS for individual atomic species. Notably, the dominant role of p-states becomes evident. B-p and N-p states contribute substantially to forming the valence band. On the other hand, the N-p states play a pivotal role in forming the conduction band. 

\begin{figure}[!htb]
	\centering
	\includegraphics[width=\linewidth]{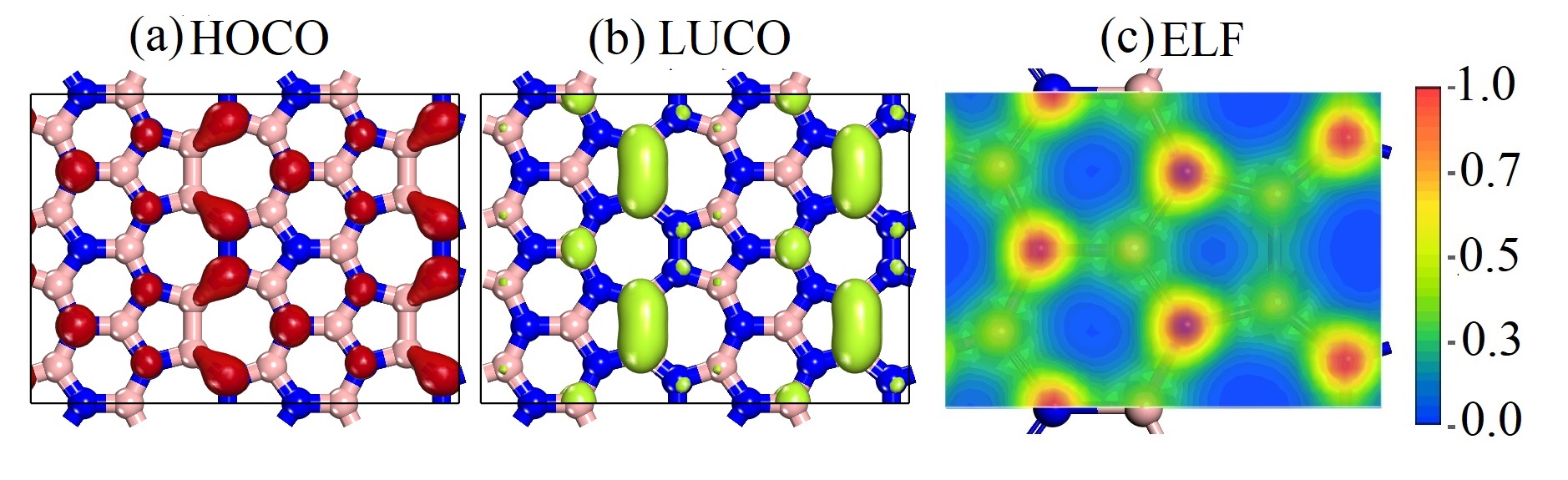}
	\caption{Schematic representation of the (a) highest occupied crystal orbital (HOCO), (b) lowest unoccupied crystal orbital (LUCO), and (c) electron localization function. These properties were calculated using the HSE06 approach.}
	\label{fig-elf}
\end{figure}

To gain a deeper understanding of the underlying chemical interactions within $\Psi$-BN, we present the results of our analyses involving the highest occupied crystal orbital (HOCO), lowest unoccupied crystal orbital (LUCO), and electron localization function (ELF) in Figure \ref{fig-elf}.

Within $\Psi$-BN, distinctive features emerge in the localization of HOCO (Figure \ref{fig-elf}(a)) and LUCO (Figure \ref{fig-elf}(b)). The HOCO primarily localizes around the B-N bonds and nitrogen atoms. At the same time, the LUCO predominantly resides within the B-B bonds, with a minor presence in the N-N bonds and boron atoms. This arrangement creates a charge imbalance, as boron and nitrogen can carry partial negative charges. Electrons concentrate near the nitrogen atoms in the HOCO, resulting in a reversed charge distribution compared to the LUCO. This shift arises from the relatively higher energy level of the nitrogen p-orbitals compared to the boron p-orbitals.

Complementing the electron density analysis in $\Psi$-BN, the electron localization function (ELF) shown in Figure \ref{fig-elf}(c) offers valuable insights into the distribution of electrons within the material. ELF provides a topological view of electron interactions, helping to identify regions of electron localization or delocalization. It assigns a value between 0 and 1 to each point in space, where values approaching 1 generally indicate strong covalent interactions or lone pair electrons. In contrast, lower values (around $\sim$0.5) signify delocalization, ionic bonds, or weak Van der Waals interactions.

In this context, the ELF map depicted in Figure \ref{fig-elf}(c) highlights localized electrons around nitrogen atoms and delocalized electrons around boron atoms. This observation aligns with the previously discussed concept of a charge imbalance in the HOCO and LUCO, resulting from the differing electronegativities of boron and nitrogen, which lead to distinct electron concentration regions.

In $\Psi$-BN, the effective masses of holes ($m_h^*$) and electrons ($m_e^*$) are measured at 0.39$m_0$ and 0.19$m_0$, respectively, with $m_0$ representing the electron mass. Generally, a lower effective mass is associated with enhanced carrier mobility, reducing resistance as carriers move through the crystal lattice.

These findings indicate that $\Psi$-BN may excel in applications requiring efficient hole transport, such as p-type transistor channels or hole-dominant charge transport layers in photodetectors. Notably, the effective masses of electrons and holes in $\Psi$-BN are lower than those in conventional h-BN, around 0.54$m_0$ \cite{cao2013two}. This characteristic suggests that $\Psi$-BN systems are better suited for applications demanding higher hole and electron transport efficiencies.

\begin{figure}[!htb]
	\centering
	\includegraphics[width=0.45\linewidth]{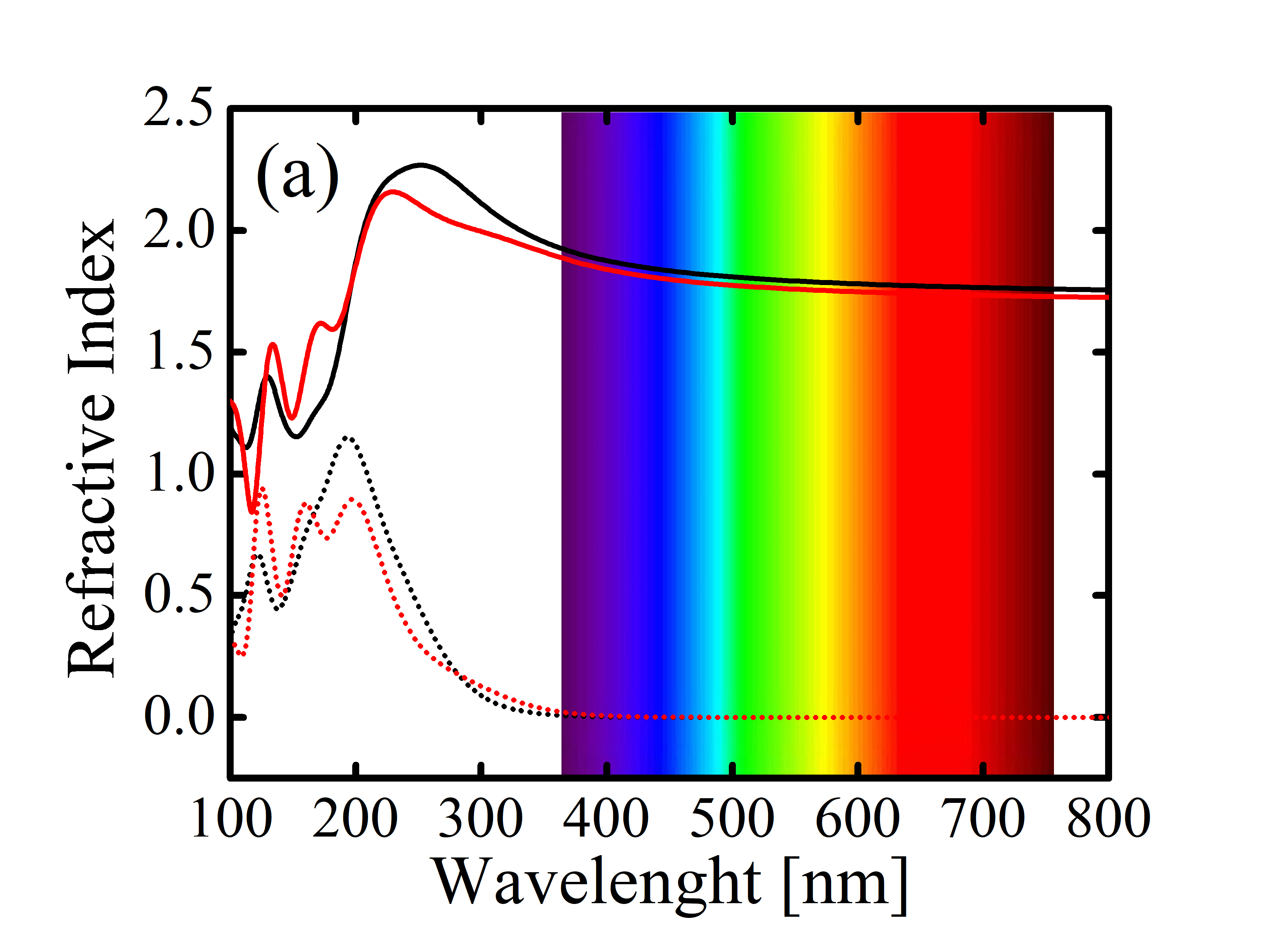}
        \includegraphics[width=0.45\linewidth]{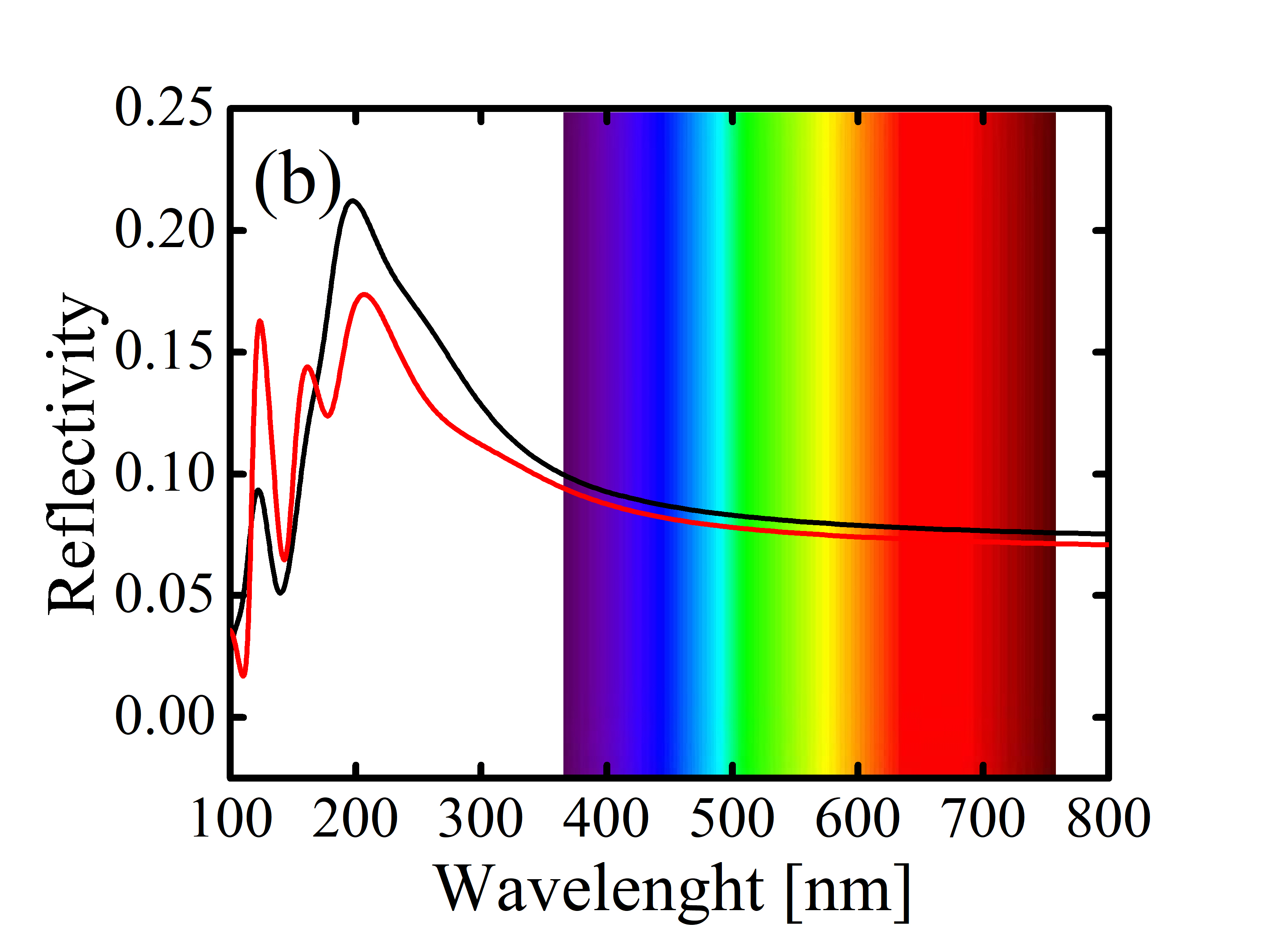}
        \includegraphics[width=0.45\linewidth]{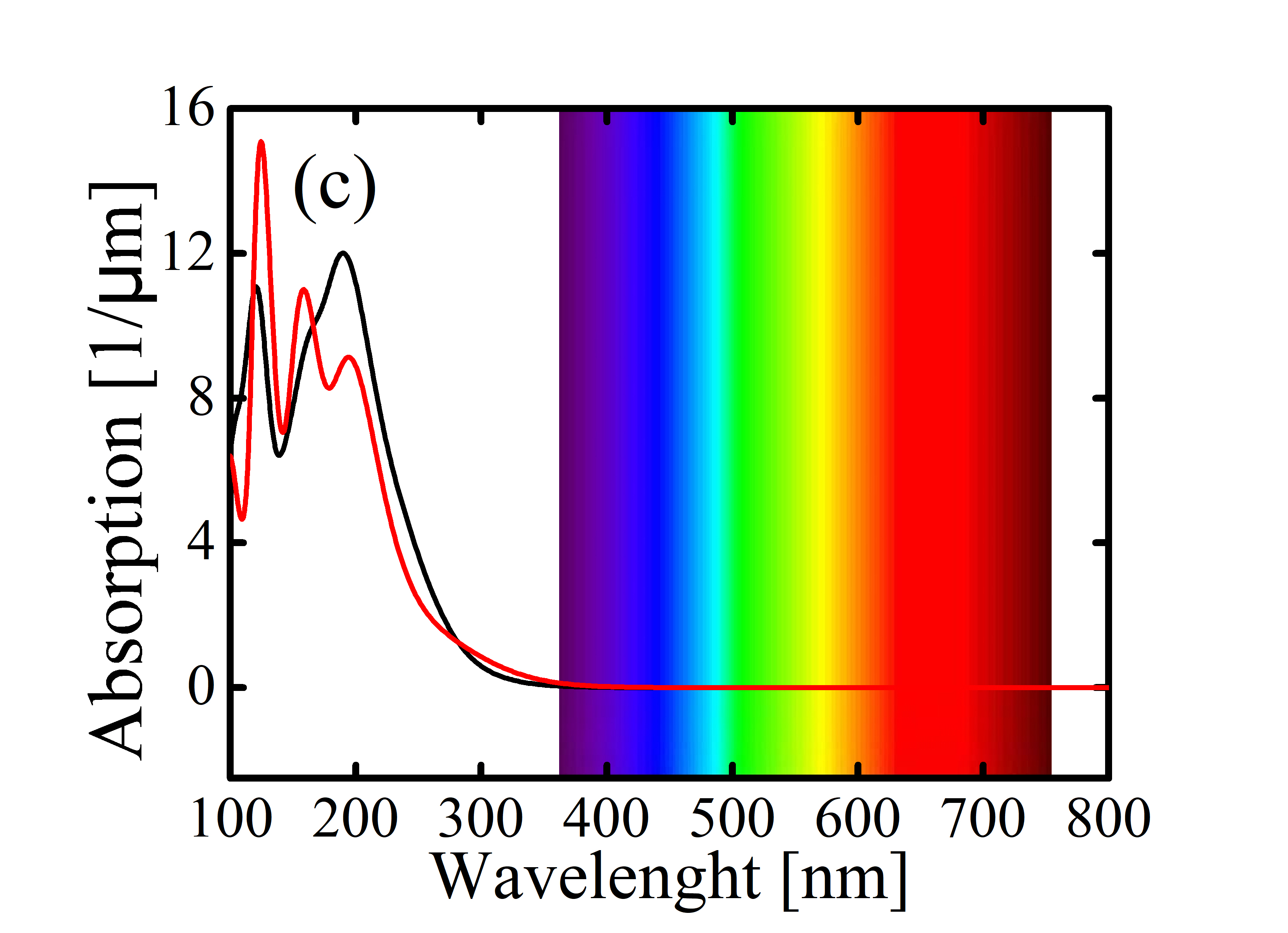}
	\caption{(a) refractive index, (b) reflectivity, and (c) absorption coefficient ($\alpha$) calculated at HSE06 level for polarised light beam oriented x (E//X) and y (E//Y) directions related to the material's surface. The solid and dashed lines correspond to the dielectric function's real (Re) and imaginary (Im) parts.}
	\label{fig-optical}
\end{figure}

Figure \ref{fig-optical} illustrates the optical properties of the $\Psi$-BN lattice, with a specific focus on the polarization of light along the x (E//X) and y (E//Y) directions. These calculations were carried out using the HSE06 method. Solid and dashed lines represent the dielectric function's real (Re) and imaginary (Im) parts in this figure.

An immediate observation from this figure is the strong optical activity exhibited by $\Psi$-BN materials, primarily in the ultraviolet (UV) region. This UV optical activity is reminiscent of h-BN, which also has a broad wavelength range, encompassing UV and visible light \cite{segura2018natural,wang2017optical}. It is worth noting that h-BN typically possesses a UV band gap, rendering it an insulator in terms of electronic conductivity \cite{cassabois2016hexagonal}. Notably, both $\Psi$-BN and h-BN owe their transparency to their wide band gaps, preventing the absorption of photons with energies below their respective band gap values.

Like h-BN, $\Psi$-BN exhibits relatively high refractive index values, as shown in Figure \ref{fig-optical}(a). This feature suggests its capability to polarize in response to external electric fields, opening up promising applications in electronics and photonics. These applications include its potential use as substrates for two-dimensional materials or protective coatings.

Interestingly, $\Psi$-BN demonstrates low reflectivity, mainly in the UV region, as depicted in Figure \ref{fig-optical}(b). Furthermore, its optical absorption is significant but limited to the UV region, as shown in Figure \ref{fig-optical}(c). These characteristics suggest that $\Psi$-BN could be an effective UV collector.

Lastly, we examine the elastic properties of $\Psi$-BN, a key aspect in comprehending microcrack behavior and materials' overall durability. To evaluate the anisotropy inherent in their mechanical characteristics, we ascertain the Poisson's ratio ($\nu(\theta)$) and Young's modulus ($Y(\theta)$) under pressure within the xy plane, as defined by the following equations \cite{doi:10.1021/acsami.9b10472,doi:10.1021/acs.jpclett.8b00616}:

\begin{equation}
    \displaystyle Y(\theta) = \frac{{C_{11}C_{22} - C_{12}^2}}{{C_{11}\alpha^4 + C_{22}\beta^4 + \left(\frac{{C_{11}C_{22} - C_{12}^2}}{{C_{44}}} - 2.0C_{12}\right)\alpha^2\beta^2}}
    \label{young}
\end{equation}

\noindent and 

\begin{equation}
    \displaystyle \nu(\theta)= \frac{{(C_{11} + C_{22} - \frac{{C_{11}C_{22} - C_{12}^2}}{{C_{44}}})\alpha^2\beta^2 - C_{12}(\alpha^4 + \beta^4)}}{{C_{11}\alpha^4 + C_{22}\beta^4 + \left(\frac{{C_{11}C_{22} - C_{12}^2}}{{C_{44}}} - 2.0C_{12}\right)\alpha^2\beta^2}}.
    \label{poisson}
\end{equation}

\begin{figure}[!htb]
	\centering
	\includegraphics[width=\linewidth]{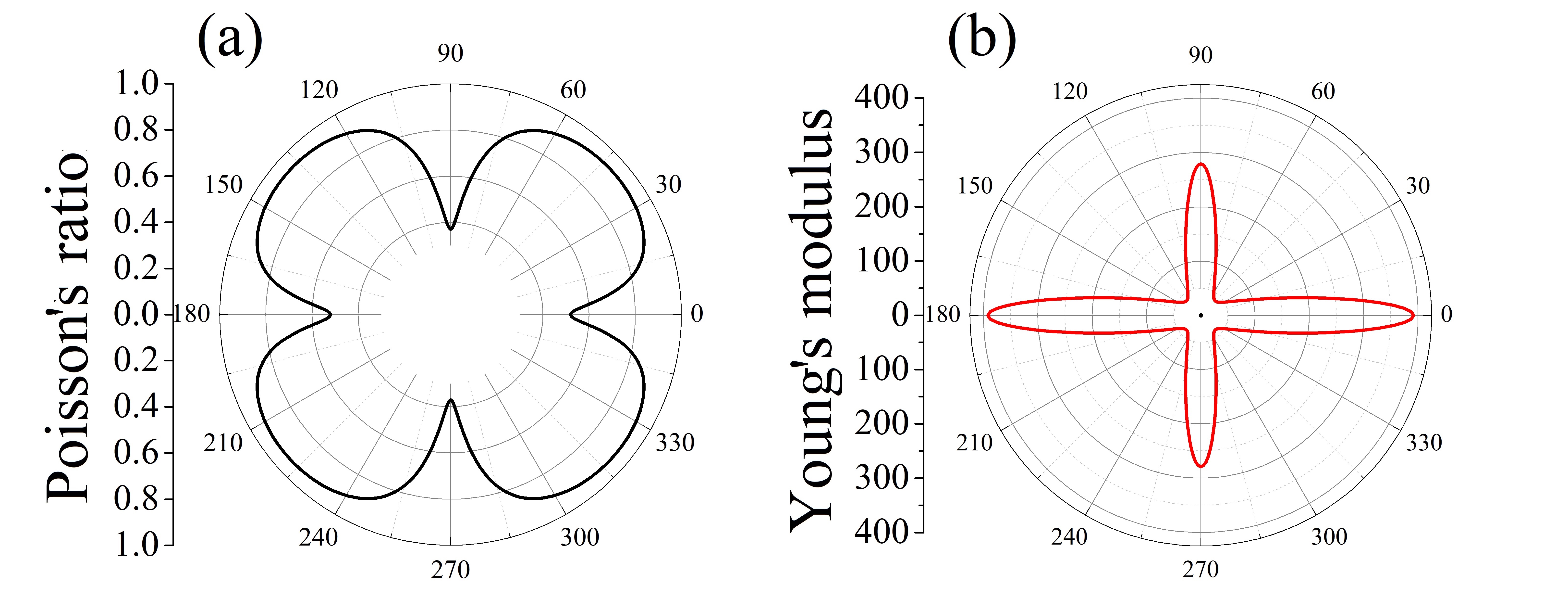}
	\caption{2D representation of (a) Poisson’s ratio and (b) Young's modulus (GPa) in the xy plane for the $\Psi$-BN monolayer.}
	\label{fig-elastic}
\end{figure}

\noindent In this context, $\alpha=\cos(\theta)$ and $\beta=\sin(\theta)$. The elastic constants characterizing $\Psi$-BN are conveniently presented in Table \ref{tab-elastic}. Additionally, Figure \ref{fig-elastic} shows the Poisson's ratio (Figure \ref{fig-elastic}(a)) and Young's modulus (Figure \ref{fig-elastic}(b)) values across the xy plane for a $\Psi$-BN monolayer.

\begin{table}[!htb]
\centering
\caption{Elastic constants C$_{ij}$ (GPa) and maximum values for Young's modulus (GPa) ($Y_{MAX}$) and maximum ($\nu_{MAX}$) and ($\nu_{MIN}$) Poisson's ratios.}
\label{tab-elastic}
\begin{tabular}{| l |c|c|c|c|c|c|c|c|}
\hline
 Structure & C$_{11}$ & C$_{12}$ &C$_{22}$ &C$_{44}$ & $Y_{MAX}$  & $\nu_{MAX}$ & $\nu_{MIN}$ \\
 \hline
$\Psi$-BN    & $345.32$       & $179.42$     & $485.04$     & $9.71$  & $  203.67$ & $0.70$ & $0.24$        \\
\hline
 \end{tabular}
\end{table}

In material mechanics, applying compressive (tensile) strain along one axis often leads to the expansion (contraction) of the material in the orthogonal direction. This phenomenon results in a positive Poisson's ratio for conventional materials. In contrast, substances exhibiting a negative Poisson's ratio are known as auxetic materials \cite{Mazaev_2020}.

As seen in Figure \ref{fig-elastic}(a), it is clear that $\Psi$-BN possesses a positive Poisson's ratio, reaching a maximum value ($\nu_{MAX}$) of approximately 0.7. Notably, for this $\nu_{MAX}$ magnitude, the associated Young's modulus remains below 50 GPa. This feature signifies the material's incompressibility when subjected to biaxial strains, encompassing the x and y directions.

As expected, $\Psi$-BN exhibits anisotropic behavior regarding Young's modulus values when subjected to strain, owing to the distinctive arrangement of its rings within its plane. The respective $Y_{MAX}$ values for strain along the x and y directions measure approximately 300 GPa and 250 GPa. It's worth noting that theoretical calculations projected Young’s modulus of 716–977 GPa (equivalent to 239–326 Nm$^{-1}$ with an effective thickness of 0.334 nm) for h-BN \cite{falin2017mechanical}.

In most common materials, Poisson's ratios typically fall within the range of 0.2 to 0.5 \cite{greaves2011poisson}. A Poisson's ratio of 0.5 indicates incompressible materials that maintain their lateral dimensions even under axial strain.

For $\Psi$-BN, the minimum Poisson's ratio values ($\nu_{MIN}$) when subjected to uniaxial strains along the x and y directions are 0.49 and 0.38, respectively. These values significantly exceed those found in h-BN, approximately 0.21 \cite{falin2017mechanical}. The heightened Poisson's ratio in $\Psi$-BN can be attributed to its unique structure, particularly the inclusion of pentagonal rings. These rings enhance rigidity compared to the hexagonal rings in h-BN, preventing $\Psi$-BN from undergoing substantial deformation under tension, resulting in lower Poisson ratios.

\section{Conclusion}

In summary, we have investigated the mechanical, electronic, and optical properties of the $\Psi$-BN lattice, a boron nitride counterpart of the intriguing $\Psi$-graphene. Using DFT calculations and AIMD simulations, we unraveled the distinctive features that characterize this novel 2D material. 

Our analysis of the lattice structure of $\Psi$-BN revealed its stability. It exhibits a unique arrangement of B-B and N-N bonds belonging to rings with five and seven atoms, while hexagons consistently present B-N bonds. Moreover, phonon calculations and AIMD simulations highlighted $\Psi$-BN's dynamic and thermal stability. We also explored the band characteristics of $\Psi$-BN. The calculated band gaps are 3.34 eV and 4.59 eV at the PBE and HSE06 levels, respectively. The effective masses of holes and electrons within $\Psi$-BN are measured at 0.39$m_0$ and 0.19$m_0$, respectively.

The optical properties of $\Psi$-BN revealed its strong ultraviolet activity, suggesting its potential as an efficient UV collector. The material's high refractive index values indicated its ability to polarize under external electric fields, presenting opportunities for electronics and photonics applications. The mechanical properties highlighted the anisotropic behavior of $\Psi$-BN, including its Poisson's ratio (0.24-0.70) and Young's modulus (250-300 GPa). 

With a wide-semiconducting bandgap and delocalized electronic states,  $\Psi$-BN could be attractive for applications in nanoelectronics and photonics. Its good thermal stability and low lattice thermal conductivity make it a candidate for thermal management solutions in microelectronics. The anisotropic mechanical properties of Psi-BN offer opportunities for reinforcing lightweight materials, while its substantial UV absorption and low reflectivity position it as an efficient UV collector for sensing and detection applications.

\section*{Acknowledgements}

This work was financed by the Conselho Nacional de Desenvolvimento Cientifico e Tecnológico (CNPq) and FAP-DF. L.A.R.J acknowledges the financial support from FAP-DF grant 00193.00000 811/2021-97, FAPDF-PRONEM grant $00193.00001247/2021-20$, and CNPq grant $302922/2021-0$. F.F.M. acknowledges the financial support from FAP-DF grant $00193.00001808/2022-71$. This study was financed in part by the Coordenação de Aperfeiçoamento de Pessoal de Nível Superior - Brasil (CAPES) - Finance Code 88887.691997/2022-00. This work used resources of the Centro Nacional de Processamento de Alto Desempenho em São Paulo (CENAPAD-SP). 

\bibliographystyle{unsrt}
\bibliography{bibliography.bib}
	
\end{document}